\documentclass[aps, prb, amsmath,amssymb,
 preprint, 
 superscriptaddress,
]{revtex4-2}

\usepackage{graphicx}
\usepackage{dcolumn}
\usepackage{bm}
\usepackage{color}
\usepackage[colorlinks=true,allcolors=blue]{hyperref}
\usepackage{subcaption}

\begin{document}

\title{High Pressure Material Phase Identification with Gaussian Kernel Neural Networks}

\author{Cameron Ryan}
\email{ryan.came@northeastern.edu}
\affiliation{University of Rochester Laboratory for Laser Energetics, 250 E River Rd, Rochester, NY 14623, USA}
\affiliation{Northeastern University, 360 Huntington Ave, Boston, MA 02115, USA}

\author{Yasantha Pasindu Hetti Kankanamalage}
\affiliation{University of Rochester Laboratory for Laser Energetics, 250 E River Rd, Rochester, NY 14623, USA}

\author{Shuai Zhang}
\email{szha@lle.rochester.edu}
\thanks{Corresponding author}
\affiliation{University of Rochester Laboratory for Laser Energetics, 250 E River Rd, Rochester, NY 14623, USA}

\begin{abstract}
Phase identification is a computational task involving the classification of material structure using the positions of atoms.
Using data from Molecular Dynamics (MD) simulations, our task is to identify the material phase of a set of atoms, based on how the simulations predict their positioning relative to each other. 
By leveraging existing research on mechanistic neural network theory, particularly Neural Tangent Kernels (NTK), we develop a new and efficient method for classifying simulated material samples.
Our method involves defining a set of ``atom detectors'' that are given positions around the origin in cartesian coordinates. The sum of Gaussians centered at each atom is calculated at each atom detector, and is used as the input to a neural network.
We present theoretical and experimental evidence to show that this method is highly effective and efficient at the task. 
Our method obtains better or similar accuracy scores to commonly used methods while having a major computational speedup.
\end{abstract}

\maketitle

\section{Introduction}
Phase identification is a computational task of identifying a specific type of material structure, given the spatial arrangement of its component atoms. Adopting a formalism similar to the one used by Lazar et al. in \cite{lazar_topological_phase_identification}, we use $\mathcal{C}(n)$ to denote the space of all configurations of the $n$ nearest neighbors to any particular atom,
\begin{equation}
\mathcal{C}(n)=  \{(r_1, r_2, ..., r_n):r_i \in\mathbb{R}^3\}
\label{eq:Cn}
\end{equation}
Our task is to analyze the geometry of any atomic environment in $\mathcal{C}(n)$ such that the type of material structure that the atoms form can be identified.
Such a technique is useful for the theoretical analysis of material properties, and studying natural phenomena. One such example would be phase transitions that occur in Molecular Dynamics (MD) simulations. It is vital to know the component phases in a material sample, when only the physical coordinates of each atom are known.  

Many common methods for phase identification involve the comparison of points in $\mathcal{C}(n)$ to a known ideal arrangement of atoms, and performing the classification based on a system of comparison.
The most significant of these methods would be Polyhedral Template Matching \cite{ptm}.
In a similar vein, \cite{lazar_topological_phase_identification} determines Voronoi cell topologies for the center atom in each ideal structure, and performs phase identification by comparing observed Voronoi cell topologies to those found in the ideal structures.
These methods are highly accurate and efficient for performing phase identification on simple periodic structures (FCC, BCC, HCP, etc).
Despite this, they do possess several drawbacks when they are used on larger, more complicated structures.
They are often highly specialized and adapted to the simple structures, so that adapting the algorithm to new structures would only be possible with a considerable amount of new research.
Furthermore, these techniques are only able to identify crystal structures, and have no capabilities for non crystal phases, such as amorphous or melt. Due to these drawbacks, they lose their usefulness for most important applications of structure identification, which general require an algorithm to be able to identify these non periodic or highly complex structures.

Various recent techniques have involved the usage of the artificial neural networks to perform this task. The most interesting and successful of these approaches, has been that of Erhard et al. \cite{erhard_model}. They propose the usage of Dynamic Graph Convolutional Neural Networks  (DGCNN) \cite{dgcnn} to perform phase identification. Erhard et al. take pure phase heating simulations, and extract ``local environments" (points in $\mathcal{C}(n)$) for each constituent atom. Each of these local environments comprises of a coordinate system centered at the focal atom, and the $n$ nearest neighbors to that atom. Their approach trains a DGCNN model to classify the pure phase environments. 

In this research, we develop a new machine learning based approach to the phase identification task. Compared to the DGCNN models used by Erhard, our method is significantly faster at generating predictions, while achieving similar accuracy scores.

\section{Method}
\subsection{Background}
\noindent\textbf{Kernel Regression:} Given a dataset $(X,y)= \{(\textbf{x}_i, y_i)\}_{i=1}^n$, kernel regression is a supervised machine learning algorithm \cite{kernel_regression}. Kernel regression leverages a symmetric positive semidefinite function $k(\textbf{x}_i,\textbf{x}_j)$ which can represent the similarity between two data points $\textbf{x}_i$ and $\textbf{x}_j$. Assuming that there is some latent function $f$ such that $y_i = f(\textbf{x}_i)$, kernel regression constructs an estimate $\hat{f}$ defined as: 
\begin{equation}
\hat{f}(\textbf{x}) = \sum_{i=1}^n (K^{-1}y)_i \,\, k(\textbf{x}_i,\textbf{x})
\label{eq:kernel_regression_f}
\end{equation}
where $K$ is an $n\times n$ kernel matrix whose entries are: $K_{ij}=k(\textbf{x}_i, \textbf{x}_j)$. We can think of kernel regression as an estimate of $f(\textbf{x})$ by a weighted sum of each label $y_i$ corresponding to how similar $\textbf{x}$ is to $\textbf{x}_i$.\\

\noindent\textbf{Neural Tangent Kernel:} Suppose that $h$ is a fully connected deep neural network, with parameter vector $\theta$ that has been initialized from a Gaussian distribution $\mathcal{N}$. Work by Jacot \textit{et al.} in \cite{Jacot_NTK} demonstrates that when the width of each layer in $h$ goes to infinity, and the Stochastic Gradient Descent (SGD) learning rate goes to zero, the neural network $h(\textbf{x};\theta)$ will converge to the kernel regression solution using the kernel:
\begin{equation}
k_{\text{NTK}}(\textbf{x}_i, \textbf{x}_j)=\mathbb{E}_{\theta \sim \mathcal{N}} \frac{\partial h(\textbf{x}_i;\theta)}{\partial\theta} \cdot \frac{\partial h(\textbf{x}_j;\theta)}{\partial\theta}
\label{eq:ntk_definition}
\end{equation}
This kernel function $k_{\text{NTK}}$ is called the Neural Tangent Kernel (NTK).

\subsection{Gaussian Kernel Neural Neural Network Derivation}


\begin{figure*}[t]
    \centering
    \begin{subfigure}[t]{0.32\linewidth}
        \centering
        \includegraphics[width=\linewidth]{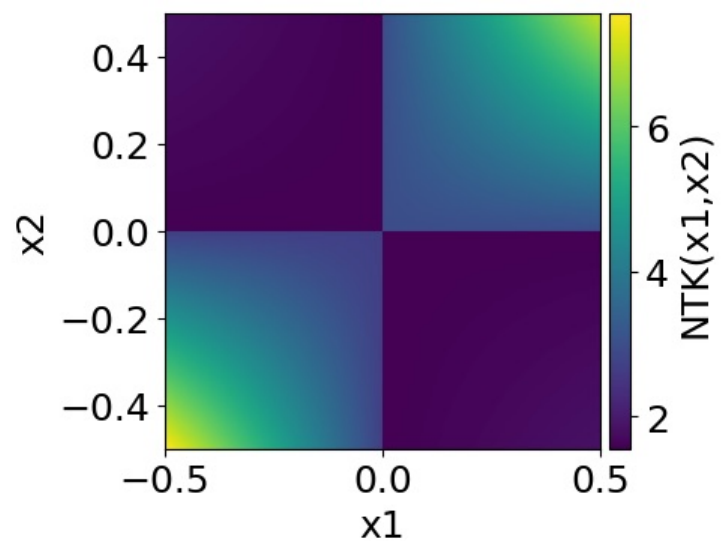}
        \caption{Raw input}
        \label{fig:a}
    \end{subfigure}
    \hfill
    \begin{subfigure}[t]{0.32\linewidth}
        \centering
        \includegraphics[width=\linewidth]{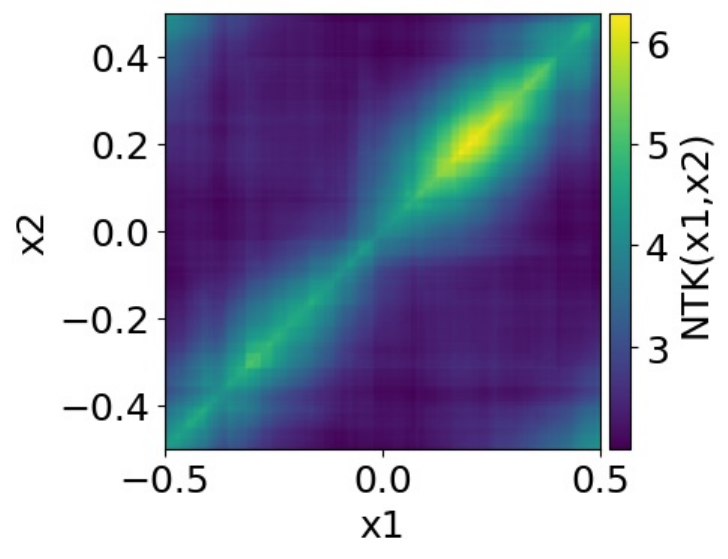}
        \caption{Fourier features}
        \label{fig:b}
    \end{subfigure}
    \hfill
    \begin{subfigure}[t]{0.32\linewidth}
        \centering
        \includegraphics[width=\linewidth]{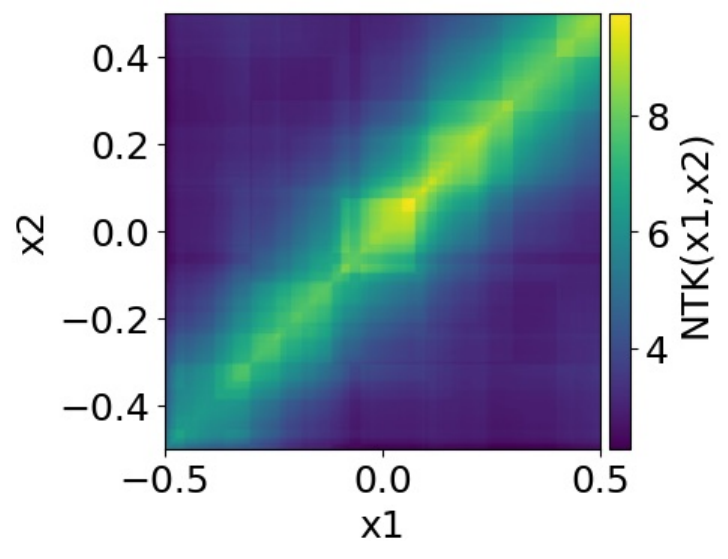}
        \caption{Gaussian kernel features}
        \label{fig:c}
    \end{subfigure}
    \caption{Neural tangent kernel of a four-layer neural network with a single input for (a) no transformation, (b) Fourier features \cite{Tancik_FF}, and (c) Gaussian kernel features for evenly distributed detectors. We see that unlike (a), (b) and (c) are able to produce stationary kernels. We adopt the method shown in (c) because of its simplicity and interpretability.}
    \label{fig:NTK_variants}
\end{figure*}

Prior work by Tancik \textit{et al.} in \cite{Tancik_FF} shows that knowledge of the NTK can be leveraged to make neural network architectures more efficient, particularly in low dimensional input domains (\textit{i.e.} mapping coordinates to classes). They show that for such problems, mapping input points to random Fourier Features will make neural networks capable of learning high frequency functions. This is because doing so causes the NTK to be more active on spatially closer points, and thus capable of judging distance. This way the NTK becomes (at the limit of the number of random features) stationary. For more information on why this is beneficial, see Appendix C of \cite{Tancik_FF}.\\
We claim that phase identification is an applicable problem. We are given atom coordinates $\textbf{r}_i=(r_1,...,r_n)\in \mathcal{C}(n)$, and we wish to classify phase from these coordinates.
The many ways that atoms can be arranged in space requires high frequency features to capture.\\ 
In a method similar to \cite{Tancik_FF}, we could compute Fourier Features for each atom such that the frequencies are drawn from the Fourier transform of the Gaussian kernel: $k(\Delta\textbf{x})=\exp(-\|\Delta \textbf{x}\|^2/2\sigma^2)$. Let us call these features $\textbf{z}(r_k)$. According to Algorithm 1 in \cite{Rahimi_random_ff}, for any point $r\in\mathbb{R}^3$ we have $\textbf{z}(r_k)^{T}\textbf{z}(r)\approx k(r_k-r)$. \\
We sum these features to ensure permutation invariance, and use these points as input to a neural network. Let $W\in \mathbb{R}^{d\times m}$ be the first weight matrix of a neural network, then the first layer activations will be: 
\begin{equation}
W\bigg(\sum_{i=1}^n \textbf{z}(r_i)\bigg)=\sum_{i=1}^{n}W\textbf{z}(r_i)
\label{eq:weight_algebra_part_1}
\end{equation}
Because of algorithm 1 in \cite{Rahimi_random_ff}, we can think of each row of the weight matrix as the frequencies for some \textit{latent} point $\tilde{r}_i$, so:
\begin{align}
&=\sum_{i=1}^n(\textbf{z}(\tilde{r}_1)^T\textbf{z}(r_i),...,\textbf{z}(\tilde{r}_m)^T\textbf{z}(r_i))^T\\
&\approx
\sum_{i=1}^n\bigg(e^{-\frac{|\tilde{r}_1-r_i|^2}{2\sigma^2}},...,e^{-\frac{|\tilde{r}_m-r_i|^2}{2\sigma^2}}\bigg)^T
\label{eq:weight_algebra_part_2}
\end{align}
Therefore, we create a set of randomly initialized points $\tilde{r}_1,...,\tilde{r}_m$, which we call the \textit{atom detectors}. We can skip the process in the first layer, and just calculate the Gaussian Kernel sum (over every atom) at each atom detector, and use that as the input to our neural network.\\
In order to understand how this is helpful, see that in Figure \ref{fig:NTK_variants} it creates a very useful NTK for distinguishing coordinates of points.
A more in depth explanation of why this allows the neural network to be efficient can be found in Appendix C of \cite{Tancik_FF}.

\subsection{Gaussian Kernel Model} 
\begin{figure}
    \centering
    \includegraphics[width=1.0\linewidth]{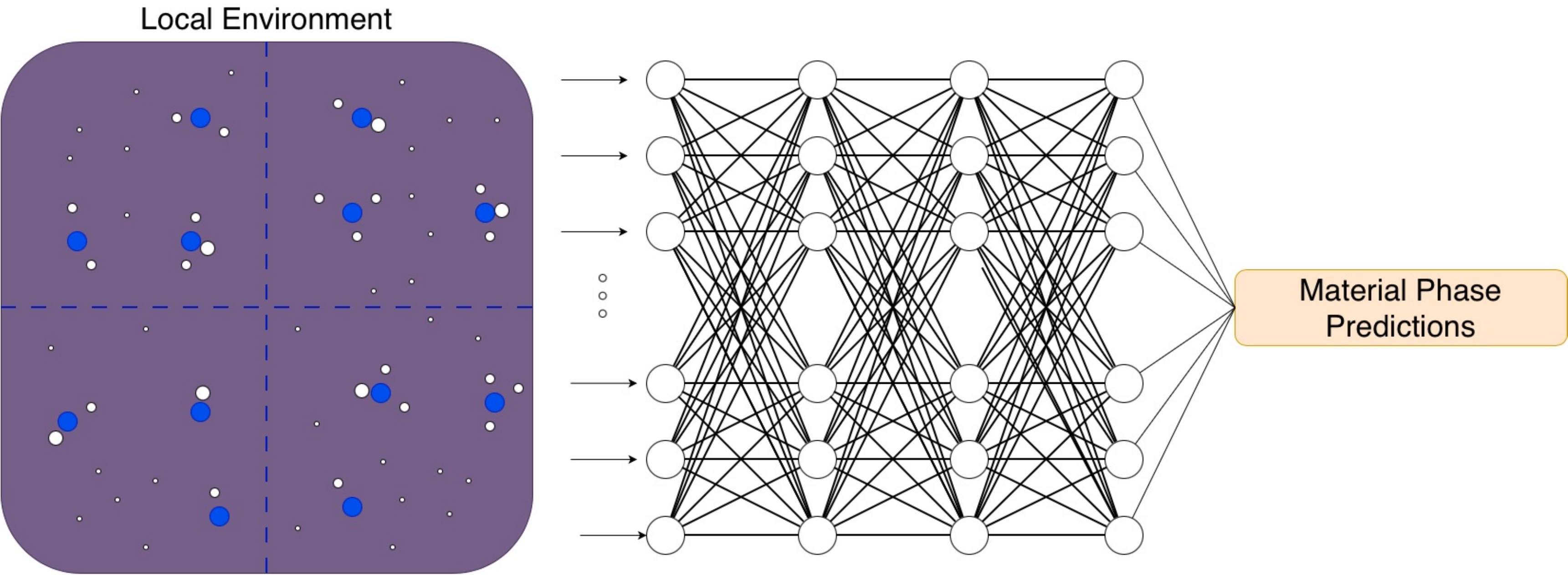}
    \caption{Schematic overview of the Gaussian Kernel model for material phase identification. Blue dots represent the coordinates of atoms in the local environment. White dots represent the positions of the ``atom detectors". The size of the white dots represents the sum of Gaussians centered at each atom at their position. Blue dashed lines indicate axes of the coordinate system. The closer that they are to an atom, the more active they are. The activation of each ``atom detector'' is the input to the neural network.}
    \label{fig:GK_model_overview}
\end{figure}
In light of our analysis of the Neural Tangent Kernel, we now have a method for using a 3 dimensional point cloud as the input to a neural network. We generate a set of atom detectors $\{\tilde{r}_1,..., \tilde{r}_m\}$ around the origin. Given some structure $\textbf{r}\in\mathcal{C}(n)$,  where $\textbf{r}=\{r_1, ..., r_n\}$, we calculate the Gaussian Kernel for a three dimensional isotropic Gaussian, at each atom detector:
\begin{equation}
g_\textbf{r}(\tilde{r}_m)=\sum_{i=1}^{n}\exp\Bigg({-\frac{|r_i-\tilde{r}_m|^2}{2\sigma^2}}\Bigg)
\label{eq:gaussian_kernel}
\end{equation}
And use the activations at each atom detector as the input to a neural network. This is visualized in Figure \ref{fig:GK_model_overview}. The coordinates of the atom detectors, and the $\sigma$ parameter are both quantities which are optimized during the training process. Experiments with initializing the atom detectors in a grid, or sampling them from a normal distribution based at the origin both yield similar results.
Our models were trained using ReLU activation functions, Adam optimizer, and cross entropy loss.
Furthermore, batches were normalized between each layer. 
\subsection{Comparison to SOAP}
Although this was derived independently, we can draw some significant parallels to Smooth Overlap of Atomic Positions (SOAP) \cite{SOAP_Bart_k_2013}. Similar to our method, SOAP represents local environments by using smooth Gaussian densities centered at each atom. However, rather than using the Gaussian densities directly, SOAP expands the density function into spherical harmonics.
Importantly, this method does provide rotational invariance.
Despite this, our experiments found that the neural networks are able to learn to output the same predictions invariant of rotation.
We hypothesize that this is due to the fact that the training process involves randomly rotating each batch of structures every epoch, and hence there exists enough data to learn rotationally invariant predictions.
Therefore, we decide not to implement this additional complexity.

\subsection{Training Data}
Training data was generated and used with the method of Erhard et al. in \cite{erhard_model}. Across various time steps, atoms were randomly selected to have their environments extracted.
For these atoms, the $n$ nearest neighbors were calculated, and a local environment was constructed with the central atom at the origin.
Furthermore, the coordinates of every other atom normalized so that the closest atom to the center has a distance of $1$ from the origin.
This way, our models can interpret structures on the same scale, independently of the environment size.
For our method this has a particular importance because atoms must be approximately located in the same region of space as the atom detectors, so that the neural network can interpret their geometry.
\\
Datasets were generated for $n=16$, $n=32$, and $n=64$ atom environment sizes. Models were trained on each dataset.

\section{Scaling Experiments}
\begin{figure}
    \centering
    \includegraphics[width=0.9\linewidth]{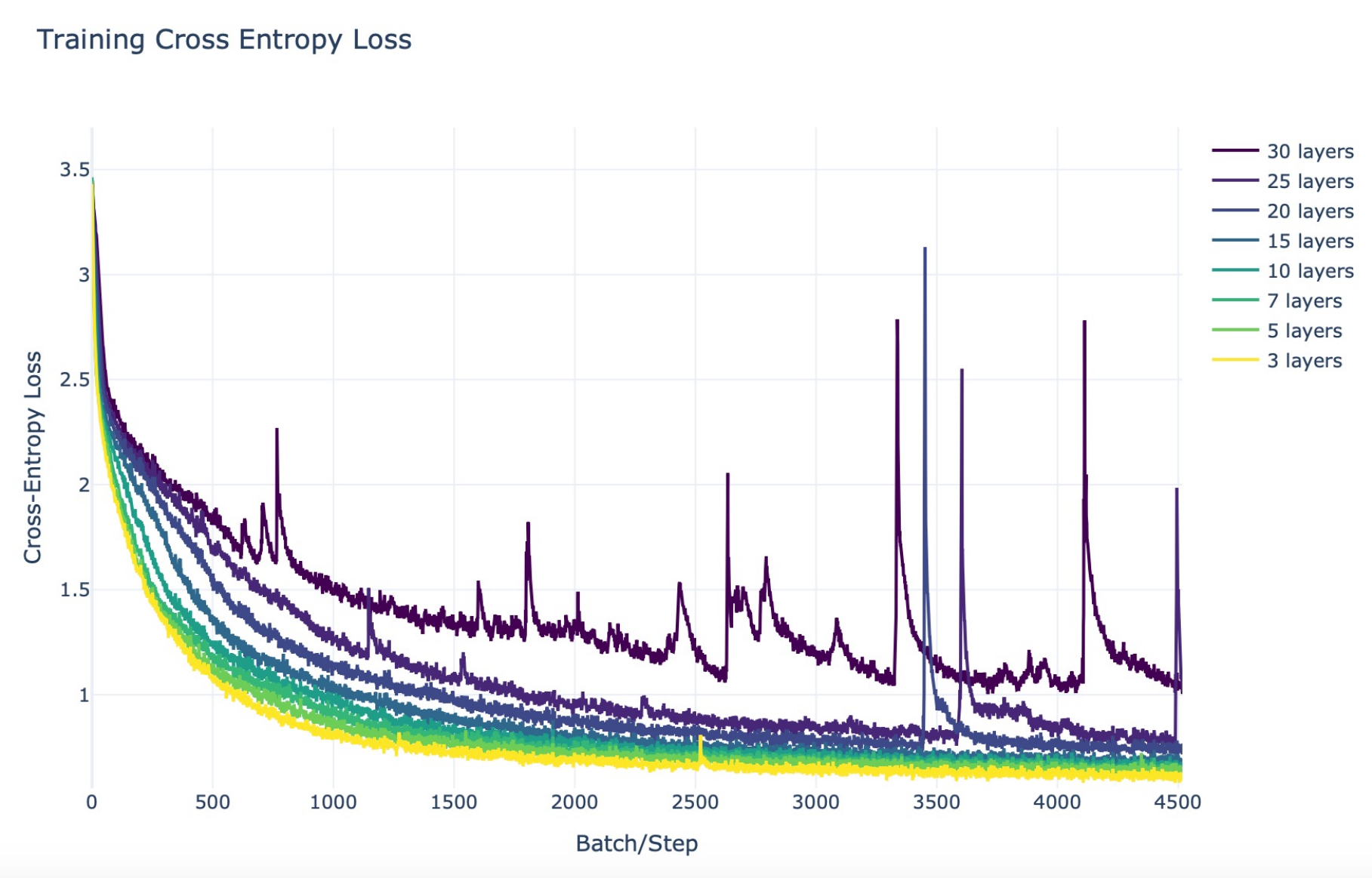}
    \caption{Loss curve during training for models with all other hyperparameters constant except for the number of layers. We find that lower number of layers allows lower loss.}
    \label{fig:layer_scaling}
\end{figure}

In our experiments, we found that the performance of these models is highly dependent upon the hyperparameters that they are initialized with. We perform a study of how the model's performance scales with number of layers, latent space dimension (number of hidden units), and learning rate. 
\begin{itemize}
    \item\textbf{Number of Layers} We find that model performance scales negatively with the number of layers. We hypothesize that this is the result of the way that the neural network learns to accomplish the task. The first two layers will detect basic symmetries and space group properties of the sample, while the last will predict a sample's phase by choosing which combinations of these properties will make up a particular phase. 
    \item\textbf{Latent Space Dimension} We find that model performance scales positively with latent space dimension. This can likely be interpreted as the model gaining more patterns which it can learn, and therefore improving its performance.
    \item\textbf{Learning Rate} We find that model performance scales negatively with learning rate. Often in loss function optimization problems, it is a common problem that having an extremely low learning rate will result in the model parameters getting stuck in relative minima, severely limiting the ability of the model to learn. This was not observed at all during the course of our experiments. We postulate that the loss landscape for this problem is highly convex, and with an extremely narrow minimum. Therefore, as the learning rate is lowered, the parameters will be able to get progressively closer to this narrow absolute minimum without being trapped in smaller relative minima.
\end{itemize}

\section{Performance Comparison}

\begin{figure}
    \centering
    \includegraphics[width=0.48\textwidth]{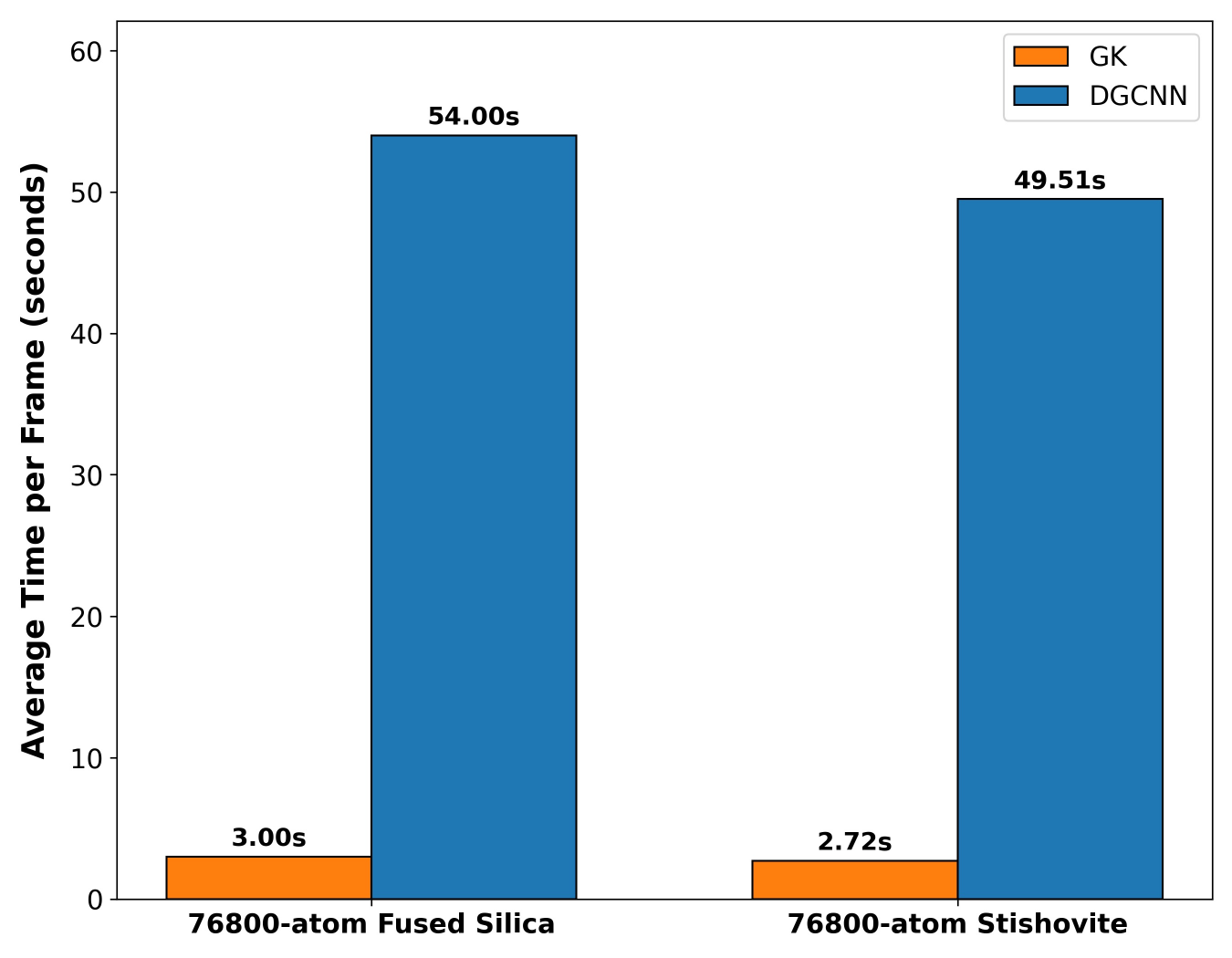}
    \caption{Comparison between the average number of seconds needed to compute predictions on a simulation frame for the Gaussian kernel and DGCNN \cite{erhard_model} models. Predictions were calculated on the same HPC environment, and done using an L40S GPU. The first simulation contains the compression of fused silica, while the second contains compression of stishovite.}
    \label{fig:computation_time_comparison}
\end{figure}

\begin{figure*}[!t]
    \centering

    \begin{subfigure}{\linewidth}
        \centering
        \includegraphics[width=\textwidth, height=0.4\textheight, keepaspectratio]{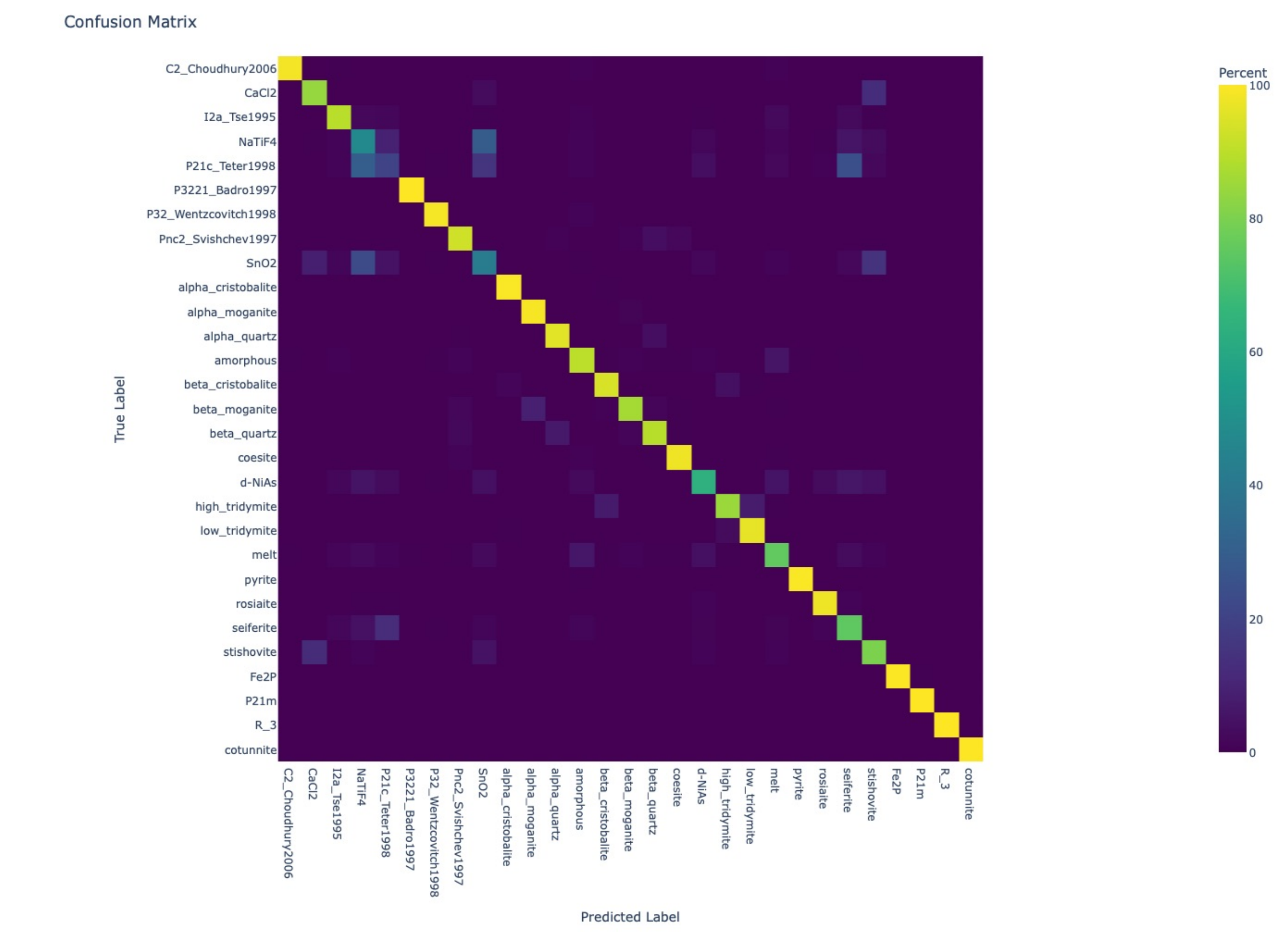}
        \caption{32 Atom Environments}
        \label{fig:top}
    \end{subfigure}

    \vspace{1em}

    \begin{subfigure}{\linewidth}
        \centering
        \includegraphics[width=\textwidth, height=0.4\textheight, keepaspectratio]{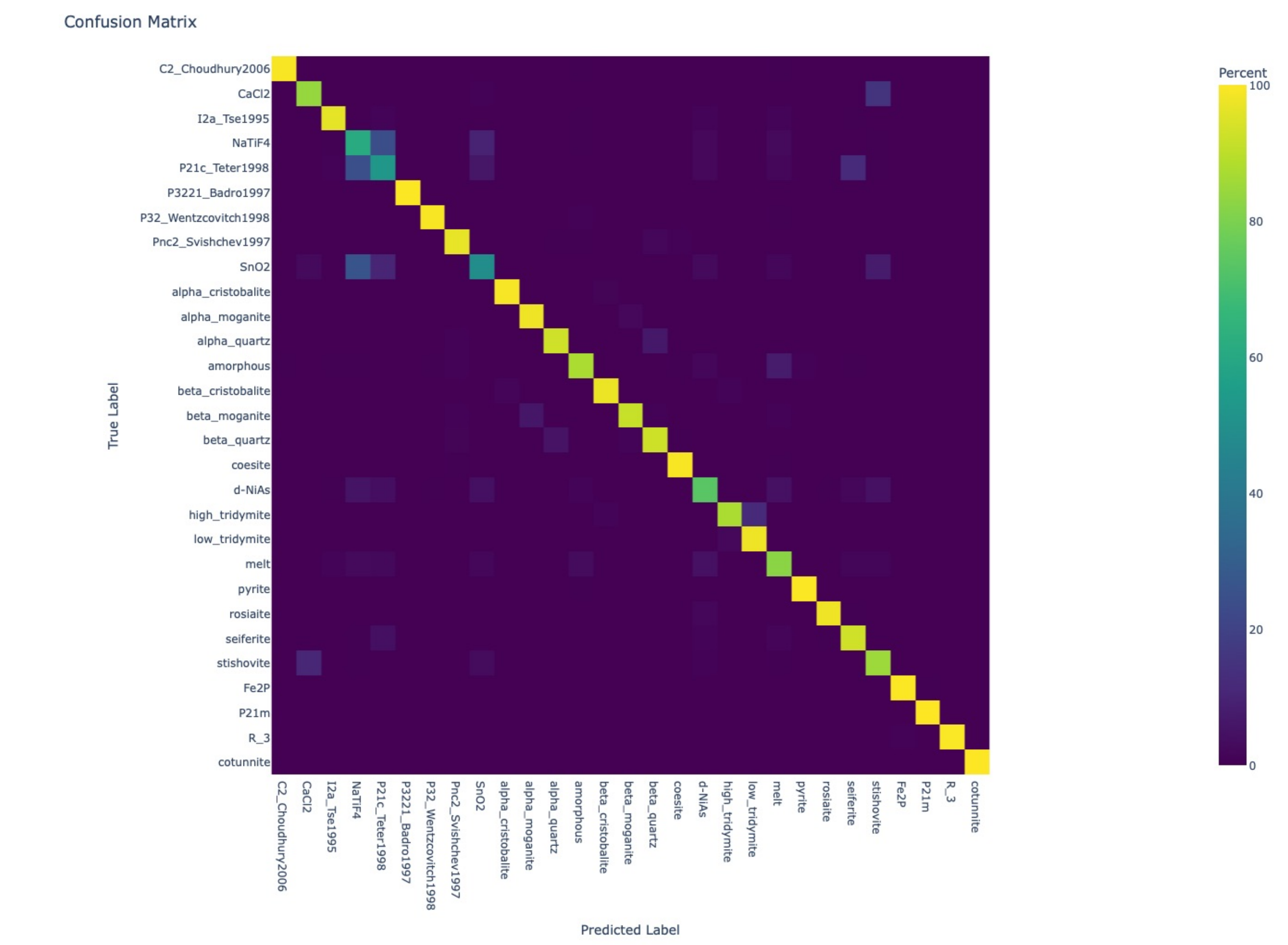}
        \caption{64 Atom Environments}
        \label{fig:bottom}
    \end{subfigure}

    \caption{Confusion matrix for the Gaussian Kernel model on $n=32$, and $n=64$ atom local environments, classifying 29 material phases.}
    \label{fig:confusion_matrices}
\end{figure*}

The Gaussian kernel model achieves an extremely large speedup compared to Erhard et al., while maintaining a similar test accuracy. 
The test accuracy scores of the Gaussian Kernel compared to Erhard et al.'s method are shown in Table \ref{tab:test_accuracy}.
Furthermore, confusion matrices for the Gaussian kernel model's predictions on our test dataset are shown in Figure \ref{fig:confusion_matrices}.
To quantify the level of speedup from this model, predictions were calculated for a simulation using the same HPC environment across different models and local environment sizes. 
By observing Figure \ref{fig:computation_time_comparison}, we see that the Gaussian Kernel model is relatively much faster, with simulation predictions being generated in around 20-30 minutes, while other models take around 5-6 hours. Each frame is computed on average about 18 times faster with than with the DGCNN.
We attribute this to the simplicity of the Gaussian Kernel Model.
The computation required for the Gaussian Kernel model only contains a few distance calculations and then a regular neural network, which is much less complicated than the DGCNN with its edge convolution operation. 

\begin{table}
\centering
\begin{tabular}{|c|c|c|}
\hline
N points & DGCNN & Gaussian Kernel \\
\hline
16 & 75\% & 78\% \\
32 & 85\% & 86\% \\
64 & 92\% & 90\% \\
\hline
\end{tabular}
\caption{Test classification accuracy for $\text{Si}\text{O}_2$ environments with 29 structure types.}
\label{tab:test_accuracy}
\end{table}

\section{Simulation Results}
\begin{figure*}[!t]
    \centering
    \includegraphics[width=1.0\linewidth]{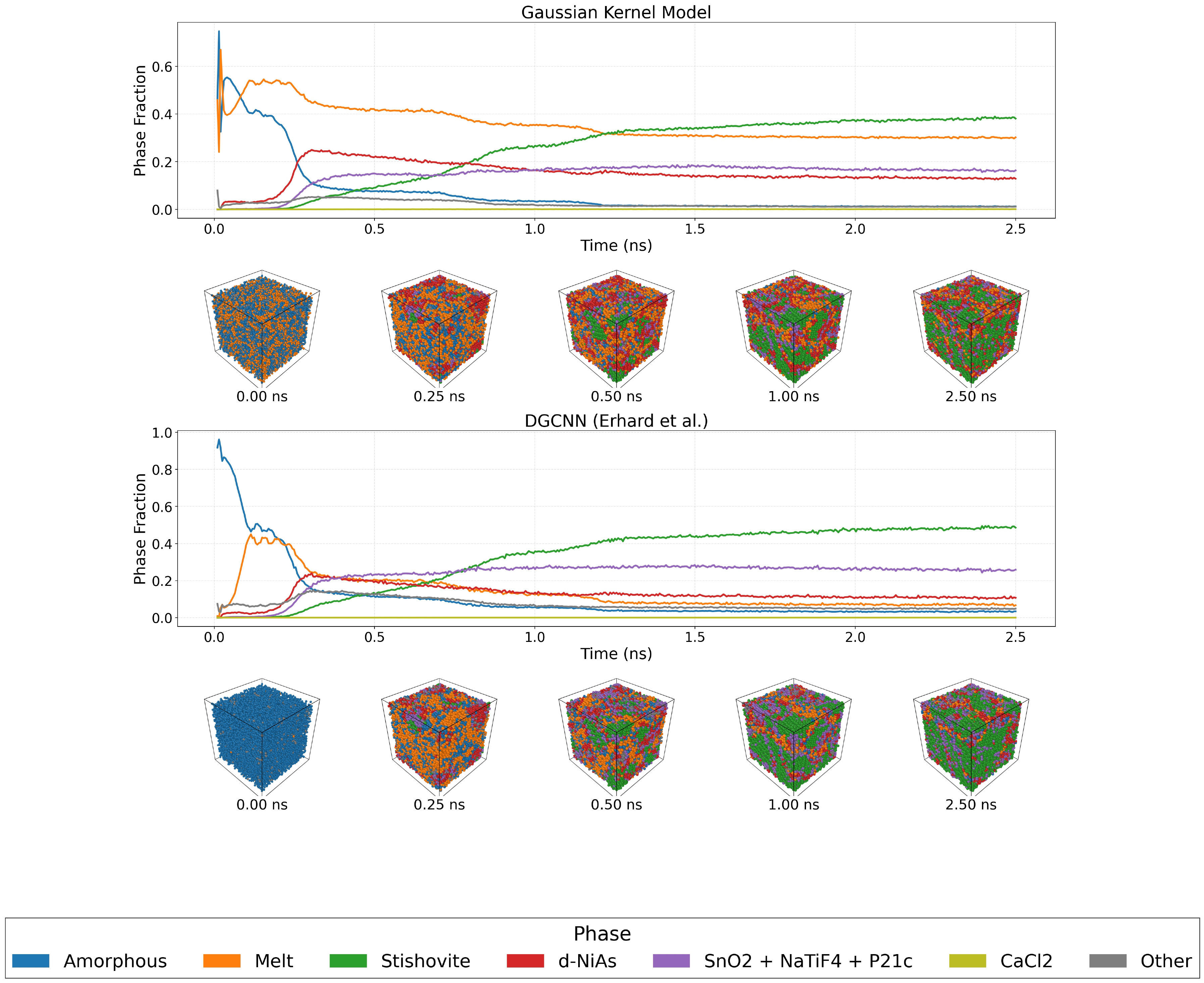}
    \caption{Predicted phase fractions on simulated shock of amorphous $\text{Si}\text{O}_2$ by a DGCNN and Gaussian kernel model with local environment size $n=64$. As expected, we see a shift from melt/amorphous to stishovite becoming the dominant phase fraction.}
    \label{fig:amorphous_sio2_shock}
\end{figure*}

\begin{figure*}[!t]
    \centering
    \includegraphics[width=1.0\linewidth]{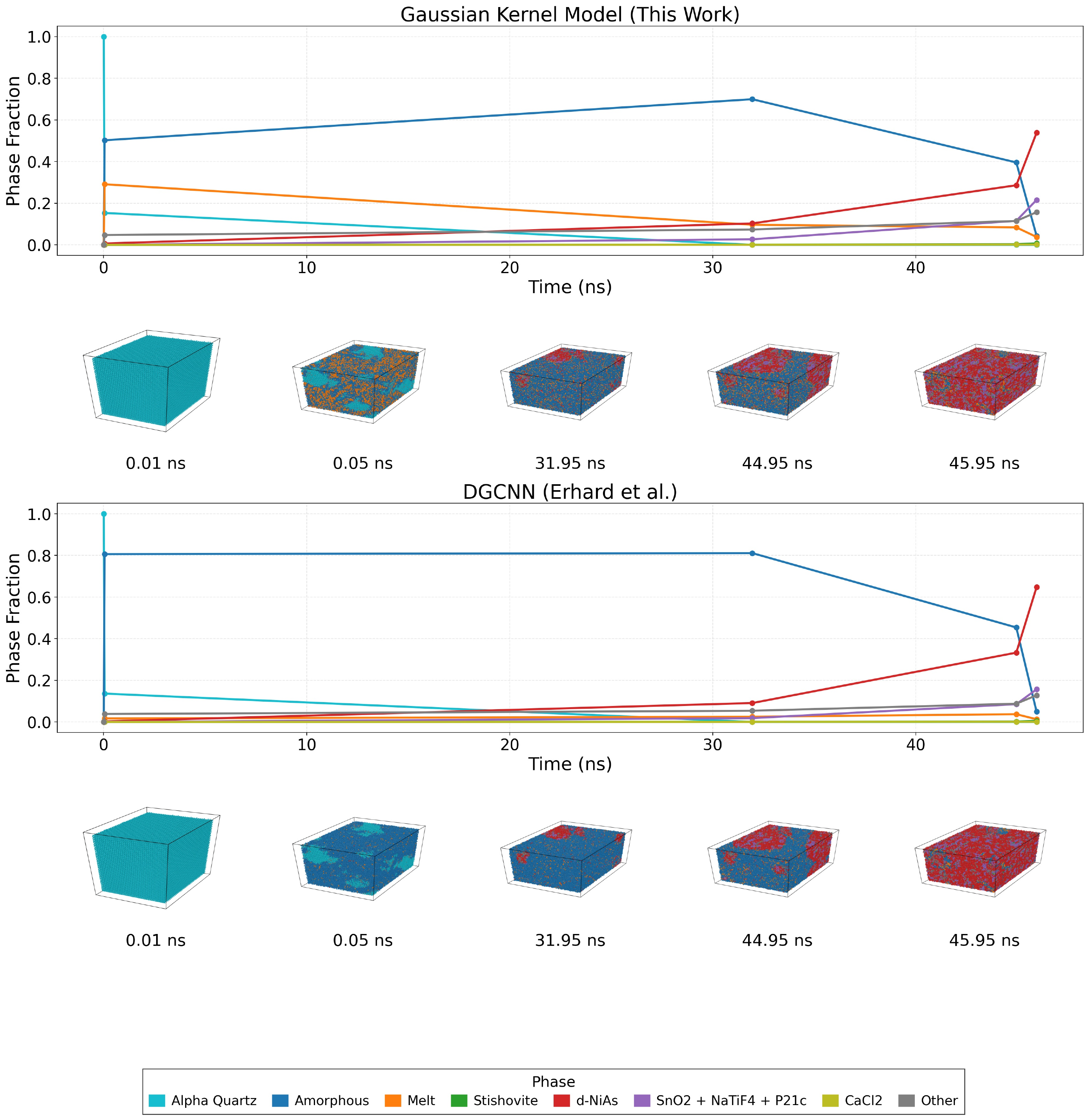}
    \caption{Predicted phase fractions on a simulated shock of alpha quartz by a DGCNN and Gaussian Kernel model. Note that the simulation involves the simulation box experiencing compression.}
    \label{fig:alpha_quartz_shock}
\end{figure*}

In order to validate the usefulness of this method, we compare its predictions on various simulations to that of \cite{erhard_model}.
On multiple molecular dynamics simulations, we extracted a local environment for every atom, and calculated the environment's phase using the Gaussian kernel model.
Then we calculate the fraction for each phase as the number of atoms predicted to be in that phase divided by the total number of atoms.
Often we can qualitatively observe that the evolution of the phase fraction curves are similar to those predicted by \cite{erhard_model}.
\\
\noindent\textbf{Fused Silica Shock:} In this simulation, an amorphous $\text{Si}\text{O}_2$ sample is subjected to shock, a process in nature that is known to cause it to undergo a transition to stishovite.
In \cite{erhard_model}, they find that their method is able to observe this phase transition occur, as the dominant phase transitions from amorphous to stishovite. 
We observe a similar result predicted by the Gaussian Kernel model, as seen in Figure \ref{fig:amorphous_sio2_shock}.
\\
\noindent\textbf{Alpha Quartz Shock:} We analyze the simulated shock of an alpha quartz sample from \cite{erhard_alpha_quartz}. In Figure \ref{fig:alpha_quartz_shock} we observe qualitatively a similarity between the predictions of the GK model and the DGCNN model. We observe that the initial phase is alpha quartz, which transitions to amorphous at the onset of shock, and later transitions to d-NiAs at the end. 

\section{Discussion}
We have proposed a method for material phase identification that is faster than existing methods while maintaining better or similar accuracies. 
We postulate that the speedup of our method is the result of its simplicity. Our method only requires distance calculations and a standard neural network, which is more straightforward than edge convolutions used by DGCNNs.
While our method is not inherently rotationally invariant due to not directly using spherical harmonics (unlike methods such as \cite{SOAP_Bart_k_2013}, or \cite{Batzner2022_E3_equivarient}), we find that the neural networks automatically learn to be rotationally invariant. We posulate that this is because the neural network learns to represent spherical harmonic features in the first layer, and therefore internally possesses a nearly rotationally invariant representation of local environments. 
We found that our models were often highly sensitive to learning rate, layers, and hidden units. Our final models were trained with 3 layers, the highest possible number of hidden units, and the lowest possible learning rate given the available computational resources. This resulted in using 4096 hidden units, and a learning rate of $5\times10^{-5}$.
The models were able to perform highly accurately on the additional high pressure phases (Fe2P, P21m, etc.).

\section{Conclusion}
The Gaussian Kernel model offers a fast and highly accurate method for performing material phase identification from atomic environment structures.
This method generates representations of local environments by randomly sampling Gaussian densities centered at each atom, and using the representations as input to a neural network.
We find that this method obtains superior or similar accuracies to that of the leading machine learning method (DGCNNs), while obtaining a massive computational speedup. 
At scale, this enables analysis of molecular dynamics simulations to be completed faster, improving the rate of research and discovery in many fields.

\section{Acknowledgments}
This material is based on work supported by the Department of Energy [National Nuclear Security Administration] University of Rochester "National Inertial Confinement Program" under award number(s) DE-NA000414. C.R. is partially supported by the U.S. DOE Office of Science under Award DE-SC0024863 (IFE-COLoR).


\begin{thebibliography}{11}%
\makeatletter
\providecommand \@ifxundefined [1]{%
 \@ifx{#1\undefined}
}%
\providecommand \@ifnum [1]{%
 \ifnum #1\expandafter \@firstoftwo
 \else \expandafter \@secondoftwo
 \fi
}%
\providecommand \@ifx [1]{%
 \ifx #1\expandafter \@firstoftwo
 \else \expandafter \@secondoftwo
 \fi
}%
\providecommand \natexlab [1]{#1}%
\providecommand \enquote  [1]{``#1''}%
\providecommand \bibnamefont  [1]{#1}%
\providecommand \bibfnamefont [1]{#1}%
\providecommand \citenamefont [1]{#1}%
\providecommand \href@noop [0]{\@secondoftwo}%
\providecommand \href [0]{\begingroup \@sanitize@url \@href}%
\providecommand \@href[1]{\@@startlink{#1}\@@href}%
\providecommand \@@href[1]{\endgroup#1\@@endlink}%
\providecommand \@sanitize@url [0]{\catcode `\\12\catcode `\$12\catcode `\&12\catcode `\#12\catcode `\^12\catcode `\_12\catcode `\%12\relax}%
\providecommand \@@startlink[1]{}%
\providecommand \@@endlink[0]{}%
\providecommand \url  [0]{\begingroup\@sanitize@url \@url }%
\providecommand \@url [1]{\endgroup\@href {#1}{\urlprefix }}%
\providecommand \urlprefix  [0]{URL }%
\providecommand \Eprint [0]{\href }%
\providecommand \doibase [0]{https://doi.org/}%
\providecommand \selectlanguage [0]{\@gobble}%
\providecommand \bibinfo  [0]{\@secondoftwo}%
\providecommand \bibfield  [0]{\@secondoftwo}%
\providecommand \translation [1]{[#1]}%
\providecommand \BibitemOpen [0]{}%
\providecommand \bibitemStop [0]{}%
\providecommand \bibitemNoStop [0]{.\EOS\space}%
\providecommand \EOS [0]{\spacefactor3000\relax}%
\providecommand \BibitemShut  [1]{\csname bibitem#1\endcsname}%
\let\auto@bib@innerbib\@empty
\bibitem [{\citenamefont {Lazar}\ \emph {et~al.}(2015)\citenamefont {Lazar}, \citenamefont {Han},\ and\ \citenamefont {Srolovitz}}]{lazar_topological_phase_identification}%
  \BibitemOpen
  \bibfield  {author} {\bibinfo {author} {\bibfnamefont {E.~A.}\ \bibnamefont {Lazar}}, \bibinfo {author} {\bibfnamefont {J.}~\bibnamefont {Han}},\ and\ \bibinfo {author} {\bibfnamefont {D.~J.}\ \bibnamefont {Srolovitz}},\ }\bibfield  {title} {\bibinfo {title} {Topological framework for local structure analysis in condensed matter},\ }\href {https://doi.org/10.1073/pnas.1505788112} {\bibfield  {journal} {\bibinfo  {journal} {Proceedings of the National Academy of Sciences}\ }\textbf {\bibinfo {volume} {112}},\ \bibinfo {pages} {E5769} (\bibinfo {year} {2015})},\ \Eprint {https://arxiv.org/abs/https://www.pnas.org/doi/pdf/10.1073/pnas.1505788112} {https://www.pnas.org/doi/pdf/10.1073/pnas.1505788112} \BibitemShut {NoStop}%
\bibitem [{\citenamefont {Larsen}\ \emph {et~al.}(2016)\citenamefont {Larsen}, \citenamefont {Schmidt},\ and\ \citenamefont {Schiøtz}}]{ptm}%
  \BibitemOpen
  \bibfield  {author} {\bibinfo {author} {\bibfnamefont {P.~M.}\ \bibnamefont {Larsen}}, \bibinfo {author} {\bibfnamefont {S.}~\bibnamefont {Schmidt}},\ and\ \bibinfo {author} {\bibfnamefont {J.}~\bibnamefont {Schiøtz}},\ }\bibfield  {title} {\bibinfo {title} {Robust structural identification via polyhedral template matching},\ }\href {https://doi.org/10.1088/0965-0393/24/5/055007} {\bibfield  {journal} {\bibinfo  {journal} {Modelling and Simulation in Materials Science and Engineering}\ }\textbf {\bibinfo {volume} {24}},\ \bibinfo {pages} {055007} (\bibinfo {year} {2016})}\BibitemShut {NoStop}%
\bibitem [{\citenamefont {Erhard}\ \emph {et~al.}(2024)\citenamefont {Erhard}, \citenamefont {Utt}, \citenamefont {Klomp},\ and\ \citenamefont {Albe}}]{erhard_model}%
  \BibitemOpen
  \bibfield  {author} {\bibinfo {author} {\bibfnamefont {L.~C.}\ \bibnamefont {Erhard}}, \bibinfo {author} {\bibfnamefont {D.}~\bibnamefont {Utt}}, \bibinfo {author} {\bibfnamefont {A.~J.}\ \bibnamefont {Klomp}},\ and\ \bibinfo {author} {\bibfnamefont {K.}~\bibnamefont {Albe}},\ }\bibfield  {title} {\bibinfo {title} {Crystal structure identification with 3d convolutional neural networks with application to high-pressure phase transitions in sio2},\ }\href {https://doi.org/10.1088/1361-651X/ad64f3} {\bibfield  {journal} {\bibinfo  {journal} {Modelling and Simulation in Materials Science and Engineering}\ }\textbf {\bibinfo {volume} {32}},\ \bibinfo {pages} {065029} (\bibinfo {year} {2024})}\BibitemShut {NoStop}%
\bibitem [{\citenamefont {Wang}\ \emph {et~al.}(2019)\citenamefont {Wang}, \citenamefont {Sun}, \citenamefont {Liu}, \citenamefont {Sarma}, \citenamefont {Bronstein},\ and\ \citenamefont {Solomon}}]{dgcnn}%
  \BibitemOpen
  \bibfield  {author} {\bibinfo {author} {\bibfnamefont {Y.}~\bibnamefont {Wang}}, \bibinfo {author} {\bibfnamefont {Y.}~\bibnamefont {Sun}}, \bibinfo {author} {\bibfnamefont {Z.}~\bibnamefont {Liu}}, \bibinfo {author} {\bibfnamefont {S.~E.}\ \bibnamefont {Sarma}}, \bibinfo {author} {\bibfnamefont {M.~M.}\ \bibnamefont {Bronstein}},\ and\ \bibinfo {author} {\bibfnamefont {J.~M.}\ \bibnamefont {Solomon}},\ }\bibfield  {title} {\bibinfo {title} {Dynamic graph cnn for learning on point clouds},\ }\href@noop {} {\bibfield  {journal} {\bibinfo  {journal} {ACM Transactions on Graphics (TOG)}\ } (\bibinfo {year} {2019})}\BibitemShut {NoStop}%
\bibitem [{\citenamefont {Wainwright}(2019)}]{kernel_regression}%
  \BibitemOpen
  \bibfield  {author} {\bibinfo {author} {\bibfnamefont {M.~J.}\ \bibnamefont {Wainwright}},\ }\bibinfo {title} {Reproducing kernel hilbert spaces},\ in\ \href@noop {} {\emph {\bibinfo {booktitle} {High-Dimensional Statistics: A Non-Asymptotic Viewpoint}}},\ \bibinfo {series and number} {Cambridge Series in Statistical and Probabilistic Mathematics}\ (\bibinfo  {publisher} {Cambridge University Press},\ \bibinfo {year} {2019})\ p.\ \bibinfo {pages} {383–415}\BibitemShut {NoStop}%
\bibitem [{\citenamefont {Jacot}\ \emph {et~al.}(2018)\citenamefont {Jacot}, \citenamefont {Gabriel},\ and\ \citenamefont {Hongler}}]{Jacot_NTK}%
  \BibitemOpen
  \bibfield  {author} {\bibinfo {author} {\bibfnamefont {A.}~\bibnamefont {Jacot}}, \bibinfo {author} {\bibfnamefont {F.}~\bibnamefont {Gabriel}},\ and\ \bibinfo {author} {\bibfnamefont {C.}~\bibnamefont {Hongler}},\ }\bibfield  {title} {\bibinfo {title} {Neural tangent kernel: convergence and generalization in neural networks},\ }in\ \href@noop {} {\emph {\bibinfo {booktitle} {Proceedings of the 32nd International Conference on Neural Information Processing Systems}}},\ \bibinfo {series and number} {NIPS'18}\ (\bibinfo  {publisher} {Curran Associates Inc.},\ \bibinfo {address} {Red Hook, NY, USA},\ \bibinfo {year} {2018})\ p.\ \bibinfo {pages} {8580–8589}\BibitemShut {NoStop}%
\bibitem [{\citenamefont {Tancik}\ \emph {et~al.}(2020)\citenamefont {Tancik}, \citenamefont {Srinivasan}, \citenamefont {Mildenhall}, \citenamefont {Fridovich-Keil}, \citenamefont {Raghavan}, \citenamefont {Singhal}, \citenamefont {Ramamoorthi}, \citenamefont {Barron},\ and\ \citenamefont {Ng}}]{Tancik_FF}%
  \BibitemOpen
  \bibfield  {author} {\bibinfo {author} {\bibfnamefont {M.}~\bibnamefont {Tancik}}, \bibinfo {author} {\bibfnamefont {P.~P.}\ \bibnamefont {Srinivasan}}, \bibinfo {author} {\bibfnamefont {B.}~\bibnamefont {Mildenhall}}, \bibinfo {author} {\bibfnamefont {S.}~\bibnamefont {Fridovich-Keil}}, \bibinfo {author} {\bibfnamefont {N.}~\bibnamefont {Raghavan}}, \bibinfo {author} {\bibfnamefont {U.}~\bibnamefont {Singhal}}, \bibinfo {author} {\bibfnamefont {R.}~\bibnamefont {Ramamoorthi}}, \bibinfo {author} {\bibfnamefont {J.~T.}\ \bibnamefont {Barron}},\ and\ \bibinfo {author} {\bibfnamefont {R.}~\bibnamefont {Ng}},\ }\bibfield  {title} {\bibinfo {title} {Fourier features let networks learn high frequency functions in low dimensional domains},\ }\href@noop {} {\bibfield  {journal} {\bibinfo  {journal} {NeurIPS}\ } (\bibinfo {year} {2020})}\BibitemShut {NoStop}%
\bibitem [{\citenamefont {Rahimi}\ and\ \citenamefont {Recht}()}]{Rahimi_random_ff}%
  \BibitemOpen
  \bibfield  {author} {\bibinfo {author} {\bibfnamefont {A.}~\bibnamefont {Rahimi}}\ and\ \bibinfo {author} {\bibfnamefont {B.}~\bibnamefont {Recht}},\ }\bibfield  {title} {\bibinfo {title} {Random features for large-scale kernel machines},\ }in\ \href@noop {} {\emph {\bibinfo {booktitle} {Advances in Neural Information Processing Systems}}},\ \bibinfo {editor} {edited by\ \bibinfo {editor} {\bibfnamefont {J.}~\bibnamefont {Platt}}, \bibinfo {editor} {\bibfnamefont {D.}~\bibnamefont {Koller}}, \bibinfo {editor} {\bibfnamefont {Y.}~\bibnamefont {Singer}},\ and\ \bibinfo {editor} {\bibfnamefont {S.}~\bibnamefont {Roweis}}}\ (\bibinfo  {publisher} {Curran Associates, Inc.})\BibitemShut {NoStop}%
\bibitem [{\citenamefont {Bartók}\ \emph {et~al.}(2013)\citenamefont {Bartók}, \citenamefont {Kondor},\ and\ \citenamefont {Csányi}}]{SOAP_Bart_k_2013}%
  \BibitemOpen
  \bibfield  {author} {\bibinfo {author} {\bibfnamefont {A.~P.}\ \bibnamefont {Bartók}}, \bibinfo {author} {\bibfnamefont {R.}~\bibnamefont {Kondor}},\ and\ \bibinfo {author} {\bibfnamefont {G.}~\bibnamefont {Csányi}},\ }\bibfield  {title} {\bibinfo {title} {On representing chemical environments},\ }\bibfield  {journal} {\bibinfo  {journal} {Physical Review B}\ }\textbf {\bibinfo {volume} {87}},\ \href {https://doi.org/10.1103/physrevb.87.184115} {10.1103/physrevb.87.184115} (\bibinfo {year} {2013})\BibitemShut {NoStop}%
\bibitem [{\citenamefont {Erhard}\ \emph {et~al.}(2025)\citenamefont {Erhard}, \citenamefont {Otzen}, \citenamefont {Rohrer}, \citenamefont {Prescher},\ and\ \citenamefont {Albe}}]{erhard_alpha_quartz}%
  \BibitemOpen
  \bibfield  {author} {\bibinfo {author} {\bibfnamefont {L.~C.}\ \bibnamefont {Erhard}}, \bibinfo {author} {\bibfnamefont {C.}~\bibnamefont {Otzen}}, \bibinfo {author} {\bibfnamefont {J.}~\bibnamefont {Rohrer}}, \bibinfo {author} {\bibfnamefont {C.}~\bibnamefont {Prescher}},\ and\ \bibinfo {author} {\bibfnamefont {K.}~\bibnamefont {Albe}},\ }\bibfield  {title} {\bibinfo {title} {Understanding phase transitions of $\alpha$-quartz under dynamic compression conditions by machine-learning driven atomistic simulations},\ }\bibfield  {journal} {\bibinfo  {journal} {npj Computational Materials}\ }\textbf {\bibinfo {volume} {11}},\ \href {https://doi.org/10.1038/s41524-025-01542-4} {10.1038/s41524-025-01542-4} (\bibinfo {year} {2025})\BibitemShut {NoStop}%
\bibitem [{\citenamefont {S.~Batzner}(2022)}]{Batzner2022_E3_equivarient}%
  \BibitemOpen
  \bibfield  {author} {\bibinfo {author} {\bibfnamefont {L.~S. e.~a.}\ \bibnamefont {S.~Batzner}, \bibfnamefont {A.~Musaelian}},\ }\bibfield  {title} {\bibinfo {title} {E(3)-equivariant graph neural networks for data-efficient and accurate interatomic potentials.},\ }\bibfield  {journal} {\bibinfo  {journal} {Nature Communications}\ }\textbf {\bibinfo {volume} {13}},\ \href {https://doi.org/https://doi.org/10.1038/s41467-022-29939-5} {https://doi.org/10.1038/s41467-022-29939-5} (\bibinfo {year} {2022})\BibitemShut {NoStop}%
\end{thebibliography}

%

\end{document}